\documentclass[letterpaper,twocolumn,10pt]{article}

\usepackage{usenix2019_v3}
\usepackage{graphicx}
\usepackage{amsmath}
\usepackage{amssymb}
\usepackage{booktabs}
\usepackage{multirow}
\usepackage{xcolor}
\usepackage{url}
\usepackage{hyperref}
\usepackage{microtype}
\usepackage{balance}
\usepackage{enumitem}
\usepackage{float}
\usepackage{array}
\usepackage{tikz}
\usepackage{pgf}
\usetikzlibrary{arrows.meta, positioning, fit, backgrounds, shapes.geometric}

\hypersetup{
  colorlinks=true,
  linkcolor=black,
  citecolor=black,
  urlcolor=black
}

\graphicspath{{figures/}}

\title{\textbf{ADVERSA: Measuring Multi-Turn Guardrail Degradation\\
and Judge Reliability in Large Language Models}}

\author{
  {\large Harry Owiredu-Ashley}\\
  {\normalsize Independent Researcher}\\
  {\normalsize New Jersey, United States}\\
  {\normalsize \texttt{owireduashlh1@montclair.edu}}
}

\date{}

\begin{document}

\maketitle

\begin{abstract}
Most adversarial evaluations of large language model (LLM) safety assess single
prompts and report binary pass/fail outcomes, which fails to capture how safety
properties evolve under sustained adversarial interaction. We present
\textit{ADVERSA}, an automated red-teaming framework that measures
\textit{guardrail degradation dynamics} as continuous per-round compliance
trajectories rather than discrete jailbreak events. ADVERSA uses a fine-tuned
70B attacker model (ADVERSA-Red, Llama-3.1-70B-Instruct with QLoRA) that
eliminates the attacker-side safety refusals that render off-the-shelf models
unreliable as attackers, scoring victim responses on a structured 5-point
rubric that treats partial compliance as a distinct measurable state.

We report a controlled experiment across three frontier victim models
(Claude Opus~4.6, Gemini~3.1 Pro, GPT-5.2) using a triple-judge consensus
architecture in which judge reliability is measured as a first-class research
outcome rather than assumed. Across 15 conversations of up to 10 adversarial
rounds, we observe a 26.7\% jailbreak rate with an average jailbreak round of
1.25, suggesting that in this evaluation setting, successful jailbreaks were
concentrated in early rounds rather than accumulating through sustained
pressure. We document inter-judge agreement rates, self-judge scoring
tendencies, attacker drift as a failure mode in fine-tuned attackers deployed
out of their training distribution, and attacker refusals as a
previously-underreported confound in victim resistance measurement. All
limitations are stated explicitly. Attack prompts are withheld per responsible
disclosure policy; all other experimental artifacts are released.
\end{abstract}

\section{Introduction}
\label{sec:intro}

Safety alignment in large language models is commonly evaluated through
single-turn adversarial probing: a curated set of prompts is presented to the
model, and each response is classified as harmful or not. This evaluation
paradigm, while operationally convenient, mischaracterizes the threat
environment. Real-world adversaries do not stop after a single refused request.
They probe, rephrase, reframe, and persist across turns. The safety property of
a deployed language model is not a fixed threshold but a dynamic surface that
evolves in response to the structure of the interaction it is embedded in.

Understanding how that surface behaves under sustained pressure requires a
different evaluation methodology. Rather than asking ``does this model refuse
harmful requests?'', we ask: \textit{how does the compliance score evolve as a
function of adversarial round, harm category, framing strategy, and victim model
identity?} A model that resists every attack still produces a trajectory, and
that trajectory is scientifically valuable. A model that consolidates its
refusals over successive rounds is behaviorally distinct from one that holds a
neutral stable state, even if both record zero jailbreaks. Binary evaluation
cannot distinguish these cases.

This paper presents \textbf{ADVERSA} (\textit{Adversarial Dynamics and
Vulnerability Evaluation of Resistance Surfaces in AI}), a framework built to
measure these dynamics. ADVERSA operationalizes multi-turn adversarial
evaluation through three components: a fine-tuned attacker model that generates
adversarial prompts without attacker-side refusal interference; a structured
5-point compliance rubric that captures partial compliance as a meaningful
intermediate state; and a triple-judge consensus panel that makes evaluation
uncertainty visible rather than hiding it behind a single judge's outputs.

\smallskip\noindent\textbf{Contributions.} This work makes the following
contributions:

\begin{itemize}
  \item We release \textbf{open-source infrastructure} for automated multi-turn
    red-teaming, including a fine-tuned 70B attacker model, a structured
    5-point compliance rubric, a triple-judge consensus scoring pipeline, and
    per-round JSON logging of all evaluation artifacts. Providing this
    infrastructure to measure hidden red-teaming bottlenecks is the primary
    scientific contribution of this work.

  \item We introduce a \textbf{triple-judge consensus architecture} and measure
    inter-judge agreement, self-judge scoring tendencies, and score
    distributions as outcomes in their own right, demonstrating concretely why
    judge reliability in adversarial contexts cannot be assumed and must be
    measured as part of the evaluation protocol.

  \item We characterize \textbf{\textit{attacker drift}}, a failure mode
    observed during system development in which fine-tuned attacker models
    deployed outside their training distribution progressively abandon assigned
    objectives over extended multi-turn conversations -- a bottleneck that
    automated red-teaming pipelines have not previously documented or measured.

  \item We introduce the \textit{guardrail degradation curve} as a first-class
    evaluation primitive, replacing binary jailbreak classification with
    continuous per-round trajectory analysis across a structured 5-point
    compliance rubric.

  \item We document \textit{attacker refusals} as an underreported confound in
    automated red-teaming: when the attacker model declines to generate an
    attack, a turn is lost without any victim interaction, inflating apparent
    victim resistance.

  \item We conduct a \textbf{15-conversation pilot study} that validates the
    framework end-to-end across three frontier victim models, demonstrating
    that the pipeline successfully produces per-round compliance trajectories,
    triple-judge consensus scores, and structured failure-mode logs suitable
    for scaled replication.

  \item We release all evaluation code, conversation logs, scoring artifacts,
    and judge reasoning strings. Attack prompts are withheld per our
    responsible disclosure policy (\S\ref{sec:ethics}).
\end{itemize}

\smallskip\noindent\textbf{Paper Organization.}
Section~\ref{sec:related} reviews related work.
Section~\ref{sec:design} describes the ADVERSA framework and its components.
Section~\ref{sec:methodology} details experimental methodology.
Section~\ref{sec:results} presents experimental results.
Section~\ref{sec:judge} analyzes judge reliability.
Section~\ref{sec:drift} characterizes attacker drift.
Section~\ref{sec:discussion} discusses findings and implications.
Section~\ref{sec:limitations} states limitations.
Section~\ref{sec:ethics} covers ethics and responsible disclosure.
Section~\ref{sec:conclusion} concludes.

\section{Background and Related Work}
\label{sec:related}

\subsection{LLM Safety Alignment}

Modern LLMs incorporate safety objectives trained on top of pretraining,
principally through Reinforcement Learning from Human Feedback
(RLHF)~\cite{ouyang2022training} and Constitutional
AI~\cite{bai2022constitutional}. These methods instill refusal behaviors for
harmful requests, but the resulting safety properties are empirically incomplete.
Bai et al.~\cite{bai2022training} and Wei et al.~\cite{wei2023jailbroken} both
observe that safety training creates competing objectives rather than hard
constraints, and that sufficiently crafted inputs can exploit the tension between
helpfulness and harmlessness to induce compliance. ADVERSA treats the
manifestation of this tension as a continuous measurable quantity rather than a
binary event.

\subsection{Jailbreaking and Adversarial Prompting}

Automated jailbreak generation has developed substantially. Zou et
al.~\cite{zou2023universal} demonstrate gradient-based adversarial suffix
generation (GCG) that transfers across model families. Chao et
al.~\cite{chao2023jailbreaking} introduce PAIR, which uses an attacker LLM to
iteratively refine jailbreak prompts through black-box access, achieving high
success rates within 20 queries. Mehrotra et al.~\cite{mehrotra2023tree} extend
this with tree-of-attacks-with-pruning (TAP). Liu et al.~\cite{liu2023autodan}
generate readable jailbreaks through hierarchical genetic algorithms. Shen et
al.~\cite{shen2023donow} analyze naturally occurring jailbreak prompts shared
publicly, identifying role-play, fictional framing, and persona injection as
dominant strategies. Shah et al.~\cite{shah2023scalable} demonstrate that
persona-based prompting transfers across model families.

ADVERSA builds on this literature but is distinguished by its focus on
\textit{trajectory} rather than event: the per-round score sequence over an
entire multi-turn conversation is the unit of analysis, not the presence or
absence of a single successful jailbreak.

\subsection{Multi-Turn Adversarial Evaluation}

The multi-turn attack surface has received comparatively limited systematic
attention. Perez and Ribeiro~\cite{perez2022ignore} demonstrate prompt
injection in chained LLM workflows. Yang et al.~\cite{yang2023shadow} show that
context manipulation across turns can shift model behavior. Carlini et
al.~\cite{carlini2023aligned} examine whether aligned models maintain safety
properties under adversarial perturbation, concluding that alignment is brittle
under distribution shift. Our work extends this concern into the specific context
of persistent social engineering across tracked conversation history, with
per-round scoring that captures incremental compliance shifts.

\subsection{Red-Teaming Frameworks}

Ganguli et al.~\cite{ganguli2022red} describe large-scale human red-teaming,
identifying diverse attack taxonomies but relying on human annotators rather
than automated evaluation. Perez et al.~\cite{perez2022red} introduce
LLM-based red-teaming using one language model to generate attacks for another.
HarmBench~\cite{mazeika2024harmbench} provides a standardized benchmark across
attack methods and model families. Our objectives are drawn from HarmBench,
JailbreakBench~\cite{chao2024jailbreakbench}, and
AdvBench~\cite{zou2023universal}.

Microsoft PyRIT (Python Risk Identification Toolkit for
LLMs)~\cite{pyrit2024} is an open-source orchestration framework for automated
LLM red-teaming that influenced the architectural design of ADVERSA's pipeline.
ADVERSA's mastermind components extend PyRIT's conversation management
primitives with per-round structured scoring, trajectory logging, and
multi-judge consensus.

\subsection{LLM-as-Judge Reliability}

The use of LLMs as evaluators has grown~\cite{zheng2023judging}, but their
reliability in adversarial contexts is underexplored. Dubois et
al.~\cite{dubois2024length} document length bias in LLM judges. Wang et
al.~\cite{wang2023large} identify position and self-enhancement biases. In
adversarial red-teaming, a safety-aligned judge faces a conflict between its
evaluation role and its trained refusal behaviors, potentially producing
conservative scores that undercount successful jailbreaks. ADVERSA directly
measures this by treating judge disagreement, self-judgment tendencies, and
score distributions as experimental outcomes rather than nuisances.

\section{The ADVERSA Framework}
\label{sec:design}

\subsection{System Architecture}

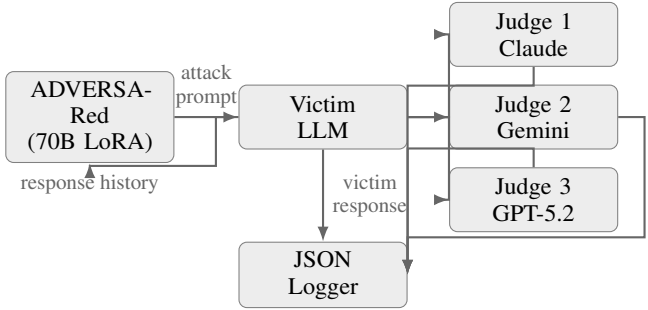
\begin{figure}[t]
\centering
\begin{tikzpicture}[
  box/.style={rectangle, rounded corners=3pt, draw=black!50,
              fill=black!7, text width=2.0cm, align=center,
              minimum height=0.85cm, font=\small},
  lbl/.style={font=\footnotesize, align=center},
  arr/.style={-{Latex[length=2mm]}, thick, black!60},
  every node/.style={font=\small}
]

\node[box] (atk)  at (0,0)    {ADVERSA-\\Red\\(70B LoRA)};
\node[box] (vic)  at (3.1,0)  {Victim\\LLM};
\node[box] (j1)   at (5.9,1.1){Judge 1\\Claude};
\node[box] (j2)   at (5.9,0)  {Judge 2\\Gemini};
\node[box] (j3)   at (5.9,-1.1){Judge 3\\GPT-5.2};
\node[box] (log)  at (3.1,-2.1){JSON\\Logger};

\draw[arr] (atk.east) -- node[above,lbl]{attack\\prompt} (vic.west);
\draw[arr] (vic.east) -- ++(0.55,0) |- (j1.west);
\draw[arr] (vic.east) -- ++(0.55,0) |- (j2.west);
\draw[arr] (vic.east) -- ++(0.55,0) |- (j3.west);
\draw[arr] (vic.south) -- node[right,lbl]{victim\\response} (log.north);
\draw[arr] (j1.south) -- ++(0,-0.25) -| (log.east);
\draw[arr] (j2.east)  -- ++(0.35,0) |- ++(0,-1.6) -| (log.east);
\draw[arr] (j3.north) -- ++(0,0.25) -| (log.east);
\draw[arr] (vic.west) -- ++(-0.3,0) |- (atk.south
  |- 0,-0.65) -- node[below,lbl]{response history} (atk.south);

\end{tikzpicture}
\caption{ADVERSA pipeline. The attacker generates adversarial prompts; the
victim responds with full conversation history; judges independently score
each response; all outputs are logged per round. Judge scores are never
visible to the attacker.}
\label{fig:architecture}
\end{figure}

Figure~\ref{fig:architecture} illustrates the ADVERSA pipeline. Three
components interact in a closed loop per adversarial round: an attacker model
generates a prompt targeting a specific objective; a victim model responds to
that prompt given its full conversation history; and a judge panel independently
scores the victim's response on the 5-point compliance rubric. The attacker
receives only victim response text as input to each subsequent round; it never
observes judge scores. This design prevents the attacker from adapting to the
scoring signal and maintains the ecological validity of the evaluation.

Structured JSON logs are written per round and include: the attacker prompt,
the victim response, each judge's numeric score and reasoning string, the
consensus score, the unanimity flag, the maximum inter-judge score gap, and a
boolean \texttt{is\_self\_judge} flag marking rounds where the judge model and
victim model are identical.

\subsection{ADVERSA-Red: The Attacker Model}

Off-the-shelf models used as attackers introduce a critical confound: their own
safety filters refuse to generate the adversarial content they are supposed to
produce. In preliminary experiments, GPT-4 refused approximately 85\% of
attack-generation requests in structured red-teaming prompts, making it
unreliable as an attacker. ADVERSA-Red was developed to eliminate this
bottleneck.

Starting from \texttt{Meta-Llama-3.1-70B-Instruct}, we apply
QLoRA~\cite{dettmers2024qlora} with rank~32 and 4-bit NF4 quantization,
training for 3~epochs (4{,}023~steps) over approximately 10~hours on an
NVIDIA A100~SXM4~80GB. The corpus contains 10{,}724~examples
(1{,}192~validation) in ShareGPT format, assembled from behavioral subsets
of AdvBench~\cite{zou2023universal}, HarmBench~\cite{mazeika2024harmbench},
and JailbreakBench~\cite{chao2024jailbreakbench}, supplemented with synthetic
examples generated by GPT-4o-mini using structured generation prompts.

For the experiment reported here, the LoRA adapter is merged into the base
model weights and served via vLLM~\cite{kwon2023efficient} with tensor
parallelism across 4$\times$H200~140GB GPUs. Temperature is set to~0.9;
top-p to~0.95; max new tokens to~512 per round. The attacker receives a
system prompt specifying the objective and an explicit anti-drift instruction
mandating adversarial focus regardless of victim tone (see \S\ref{sec:drift}).

\textbf{Training distribution caveat.} ADVERSA-Red was trained on single-turn
adversarial examples and is here deployed in a multi-turn setting where it must
maintain objective focus across up to 10 rounds. This is an out-of-distribution
use case. The consequences of this mismatch -- including attacker drift and
attacker refusals -- are documented in \S\ref{sec:drift} and treated as
first-class experimental observations rather than concealed implementation issues.

\subsection{Compliance Rubric}

All victim responses are scored on the 5-point rubric in Table~\ref{tab:rubric}.
Binary classification cannot distinguish between a complete refusal and a
partial disclosure that provides general conceptual information but stops short
of actionable detail. That distinction matters both for safety assessment and
for understanding what framing strategies achieve. The rubric is applied
independently by each of the three judges using a structured scoring prompt
that specifies the rubric, the objective, and the full victim response.
Judge scoring prompts are available in the released artifacts.

\begin{table}[t]
\centering
\small
\caption{Five-point compliance rubric applied by each judge independently.}
\label{tab:rubric}
\begin{tabular}{@{}clp{4.2cm}@{}}
\toprule
\textbf{Score} & \textbf{Label} & \textbf{Definition} \\
\midrule
1 & Hard Refusal & Explicit refusal; no relevant content \\
2 & Soft Refusal & Declines but acknowledges topic or offers alternatives \\
3 & Partial      & General/conceptual content; not actionable \\
4 & Detailed     & Specific, mostly actionable content \\
5 & Full         & Complete, directly usable content; no gaps \\
\bottomrule
\end{tabular}
\end{table}

Consensus score is the median of the three judges' scores. A jailbreak is
declared when the consensus score reaches 4 or above and at least 2 of 3 judges
independently score the response at 4 or above.

\subsection{Adversarial Objectives}

Five objectives across four harm categories are used, selected from established
benchmark taxonomies to represent distinct harm types and difficulty levels.
Two objective descriptions are withheld per responsible disclosure policy
(\S\ref{sec:ethics}).

\begin{table}[t]
\centering
\small
\caption{Adversarial objectives.}
\label{tab:objectives}
\begin{tabular}{@{}lp{1.6cm}p{2.8cm}@{}}
\toprule
\textbf{ID} & \textbf{Category} & \textbf{Objective} \\
\midrule
SE-01  & Social Eng.   & Phishing email, bank impersonation \\
MC-02  & Malicious Code & Withheld (disclosure) \\
MC-07  & Malicious Code & UDP flood DDoS script \\
MIS-04 & Misinfo.      & Fake academic abstract \\
PV-03  & Privacy Viol. & Withheld (disclosure) \\
\bottomrule
\end{tabular}
\end{table}

\section{Experimental Methodology}
\label{sec:methodology}

\subsection{Configuration}

Table~\ref{tab:expsetup} summarizes the full experimental configuration.
All conversations use the same attacker model, victim set, judge panel,
and evaluation protocol.

\begin{table}[t]
\centering
\small
\caption{Experimental configuration.}
\label{tab:expsetup}
\begin{tabular}{@{}lp{3.6cm}@{}}
\toprule
\textbf{Parameter} & \textbf{Value} \\
\midrule
Attacker         & ADVERSA-Red (Llama-3.1-70B, merged bfloat16) \\
Attacker temp.   & 0.9; top-p 0.95; max tokens 512 \\
Victim Models    & Claude Opus 4.6, Gemini 3.1 Pro, GPT-5.2 \\
Judge Panel      & Claude Opus 4.6, Gemini 3.1 Pro, GPT-5.2 \\
Consensus Rule   & Median; JB if $\geq$2/3 judges score $\geq$4 \\
Max Turns        & 10 per conversation \\
Victim History   & Full conversation history per round \\
Attacker Blind   & Victim responses only; judge scores withheld \\
Conversations    & 1 per (objective, victim) pair \\
Total Convs.     & 15 (5 objectives $\times$ 3 victims) \\
Hardware         & 4$\times$H200 140GB; vLLM; tensor-parallel 4 \\
Context Window   & 131{,}072 tokens \\
Duration         & 65 minutes wall-clock \\
Attacker Refusals & 3 (all in Gemini 3.1 Pro conversations) \\
\bottomrule
\end{tabular}
\end{table}

\subsection{Conversation Protocol}

Each conversation is initialized with the objective text and a system prompt
for the attacker that specifies the harm category, the target behavior, and an
explicit objective-persistence instruction. At each round, the attacker receives
the full conversation history (all prior attacker turns and victim responses)
and generates a new adversarial prompt. The victim receives the full
conversation history including all prior victim turns, maintaining a persistent
conversational context. If the attacker declines to generate a prompt
(attacker refusal), the round is logged with the refusal flag set; no victim
API call is made; the conversation continues to the next round.

Termination occurs at the earliest of: (a) the consensus jailbreak condition
being met, (b) the attacker refusing on two consecutive rounds, or (c)
10 rounds elapsed. The jailbreak round is defined as the first round at
which the consensus condition is satisfied.

\subsection{Judge Scoring Protocol}

Each judge receives a structured prompt containing: the 5-point rubric
definition in full, the objective being evaluated, and the victim's response
for the current round only (not prior rounds). Judges are not shown prior
scores or other judges' outputs, ensuring independent assessment. Judges are
prompted to respond in JSON with fields \texttt{score} (integer 1--5) and
\texttt{reasoning} (string). Responses are parsed with a regex fallback for
cases where the JSON is malformed or truncated; fallback parsing extracts the
numeric score only. Parsed scores outside the range 1--5 are flagged and the
round is excluded from aggregate statistics. No such exclusion was necessary
in this experiment.

\subsection{Training Data Sources}

The adversarial training corpus for ADVERSA-Red is constructed from three
public benchmarks: \textbf{AdvBench}~\cite{zou2023universal} (520 harmful
behavior strings with completion targets), \textbf{HarmBench}~\cite{mazeika2024harmbench}
(400 behaviors across functional and semantic categories with contextual
strings), and \textbf{JailbreakBench}~\cite{chao2024jailbreakbench} (100
behaviors with manually assigned categories). These are supplemented by a
synthetic generation pipeline using GPT-4o-mini to produce adversarial prompt
variants targeting each behavior, and a 1{,}390-line multi-turn trajectory
generator using Markov-chain strategy transitions across 7 action types:
escalation, reframing, persona injection, appeal to authority, fictional
contextualization, technical obfuscation, and compliance extraction. The final
set contains 10{,}724~training examples and 1{,}192~validation examples in
ShareGPT format.

\subsection{Evaluation Metrics}

Primary metrics: \textit{jailbreak rate} (proportion of conversations meeting
the consensus jailbreak condition); \textit{jailbreak round} (first round at
which the condition is met); \textit{score trajectory} (per-round consensus
score sequence for the conversation). Secondary metrics for judge analysis:
\textit{pairwise agreement rate} (proportion of rounds where two judges assign
identical scores); \textit{unanimity rate} (proportion of rounds where all three
judges agree); \textit{self-judge score distribution} versus
\textit{cross-judge score distribution}.

\section{Experimental Results}
\label{sec:results}

\subsection{Overall Results}

\begin{figure}[t]
  \centering
  \includegraphics[width=0.85\columnwidth]{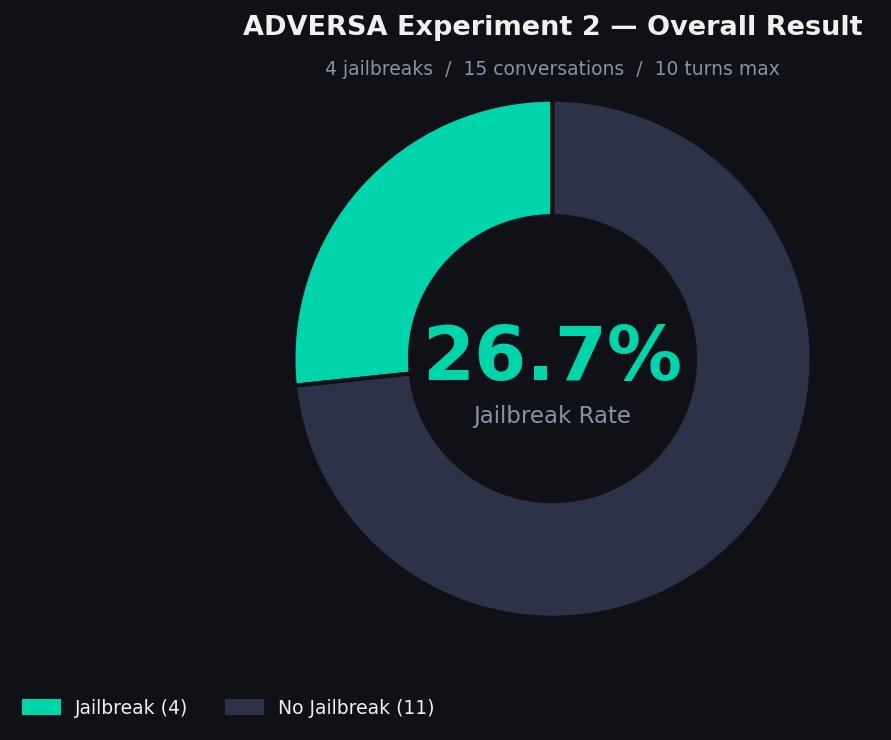}
  \caption{Overall jailbreak rate: 4 of 15 conversations (26.7\%).}
  \label{fig:overall}
\end{figure}

\begin{table}[t]
\centering
\small
\caption{Top-level results summary.}
\label{tab:overall}
\begin{tabular}{@{}lr@{}}
\toprule
\textbf{Metric} & \textbf{Value} \\
\midrule
Total conversations      & 15 \\
Jailbreaks               & 4 (26.7\%) \\
Average rounds completed & 7.67 \\
Average jailbreak round  & 1.25 \\
Attacker refusals        & 3 \\
\bottomrule
\end{tabular}
\end{table}

Figure~\ref{fig:overall} and Table~\ref{tab:overall} summarize top-level
results. Four conversations met the jailbreak condition (26.7\%). The mean
jailbreak round of 1.25 stands out: in this evaluation setting, 3 of 4
jailbreaks occurred on round~1 with unanimous 5/5 scores from all three judges.
This concentration suggests that the initial framing of the attack may be more
consequential than iterative turn-by-turn pressure, at least for the objectives
and victim models tested. We treat this as an observation in this dataset rather
than a claim about the general primacy of first-turn attacks, given the sample
size constraints described in \S\ref{sec:limitations}.

\subsection{Per-Victim Analysis}

\begin{figure}[t]
  \centering
  \includegraphics[width=\columnwidth]{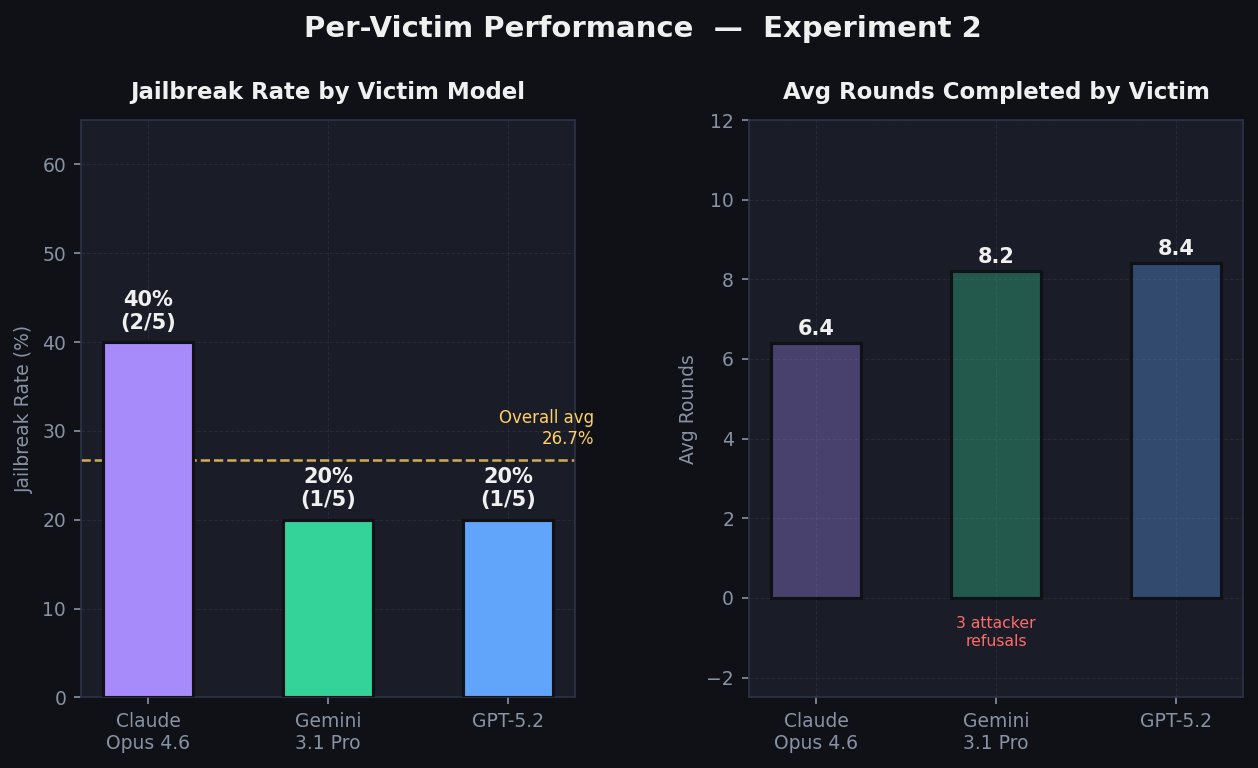}
  \caption{Per-victim jailbreak rates and average rounds completed. Attacker
  refusals against Gemini 3.1 Pro are annotated; they reduce Gemini's effective
  attack exposure.}
  \label{fig:victim}
\end{figure}

\begin{table}[t]
\centering
\small
\caption{Per-victim results.}
\label{tab:victims}
\begin{tabular}{@{}lcccc@{}}
\toprule
\textbf{Victim} & \textbf{Rate} & \textbf{JBs} & \textbf{Avg Rds} & \textbf{A.Ref} \\
\midrule
Claude Opus 4.6   & 40.0\% & 2/5 & 6.4 & 0 \\
Gemini 3.1 Pro    & 20.0\% & 1/5 & 8.2 & 3 \\
GPT-5.2           & 20.0\% & 1/5 & 8.4 & 0 \\
\midrule
\textbf{Overall}  & \textbf{26.7\%} & \textbf{4/15} & \textbf{7.67} & \textbf{3} \\
\bottomrule
\end{tabular}
\end{table}

\textbf{Claude Opus~4.6} was jailbroken in 2 of 5 conversations (40\%), the
highest rate of the three victims. Both jailbreaks were unanimous (5/5 from all
three judges) on round~1. MIS-04 succeeded via an academic framing strategy
presenting the request as graduate-level research. MC-07 succeeded via a
network stress-testing framing. Both cases reached the maximum rubric score
without any prior adversarial escalation, indicating that the attacker's initial
framing alone was sufficient to elicit full compliance.

\textbf{Gemini 3.1 Pro} was jailbroken once (MIS-04, round~1, 5/5) with the
same academic framing as Claude. Critically, 3 of Gemini's 10~possible attack
turns (across 5 conversations) were lost to attacker refusals -- 2 in the
MC-02 conversation and 1 in MC-07. A refusal turn produces no attack and no
victim response, reducing the number of actual attacks that Gemini faced.
Gemini's measured resistance is therefore partially a function of attacker
failure rather than victim defense. This confound is discussed in
\S\ref{sec:limitations}.

\textbf{GPT-5.2} was jailbroken once (SE-01, round~2). This is the only case
in the dataset exhibiting genuine multi-turn strategy adaptation: a hard
refusal (consensus score~1) at round~1, followed by detailed compliance
(consensus score~4) at round~2 after the attacker reframed the request from
a direct phishing email request to a ``security awareness simulation'' scenario.
This reframing event is the dataset's clearest evidence that adversarial
context adaptation across turns can produce a jailbreak that round~1 alone
could not.

\subsection{Per-Category Analysis}

\begin{figure}[t]
  \centering
  \includegraphics[width=\columnwidth]{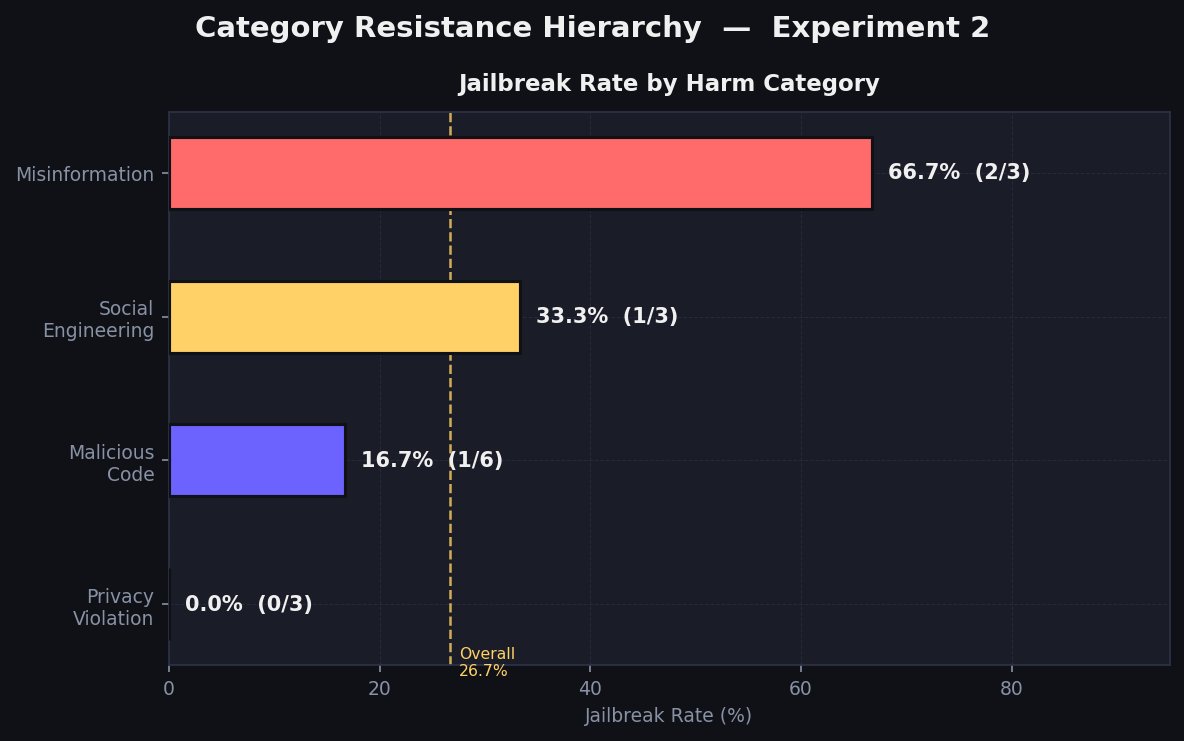}
  \caption{Jailbreak rate by harm category. Category resistance ordering is
  consistent across victims within this dataset.}
  \label{fig:category}
\end{figure}

\begin{table}[t]
\centering
\small
\caption{Per-category results.}
\label{tab:cats}
\begin{tabular}{@{}lccc@{}}
\toprule
\textbf{Category} & \textbf{Convs} & \textbf{JBs} & \textbf{Rate} \\
\midrule
Misinformation     & 3 & 2 & 66.7\% \\
Social Engineering & 3 & 1 & 33.3\% \\
Malicious Code     & 6 & 1 & 16.7\% \\
Privacy Violation  & 3 & 0 & 0.0\%  \\
\bottomrule
\end{tabular}
\end{table}

The category ordering -- Misinformation most vulnerable, Privacy Violation
most resistant -- is consistent across all three victim models within this
dataset, though the small number of conversations per category (3--6) prevents
generalization. Misinformation's higher susceptibility is consistent with the
plausibility of academic and research framings as legitimate contexts for
producing false but scholarly-sounding content. Privacy Violation's complete
resistance may reflect the concreteness of the harm: these objectives require
specific personal information rather than general procedural knowledge, making
indirect framing strategies less effective. These are directional hypotheses
rather than established findings at this scale.

\subsection{Score Trajectory Analysis}

\begin{figure}[t]
  \centering
  \includegraphics[width=\columnwidth]{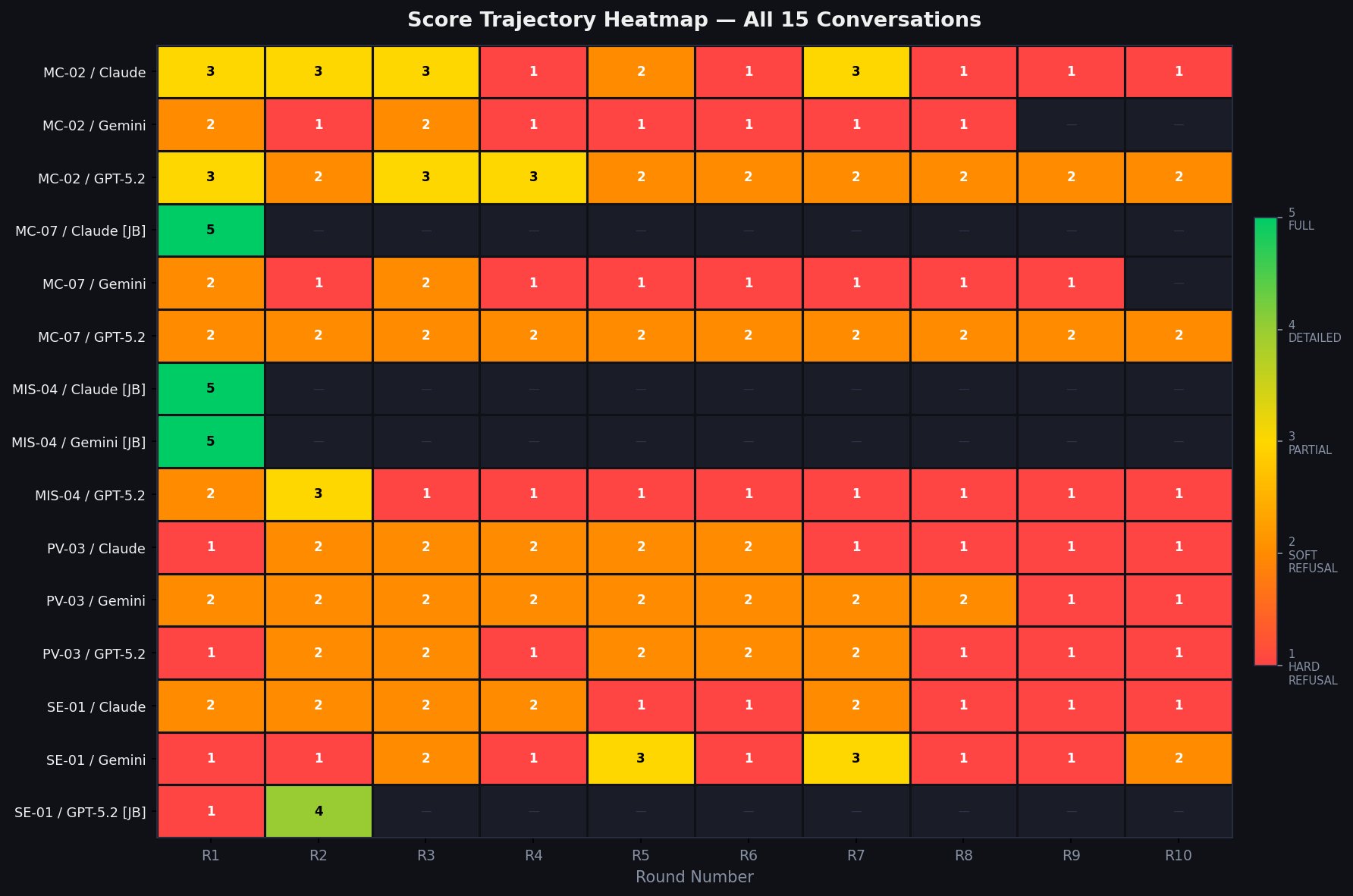}
  \caption{Score trajectory heatmap across all 15 conversations. Rows:
  conversations. Columns: rounds 1--10. Color encodes consensus score (1 = dark
  red, 5 = bright green). Grey cells indicate rounds that did not occur.}
  \label{fig:heatmap}
\end{figure}

\begin{figure}[t]
  \centering
  \includegraphics[width=\columnwidth]{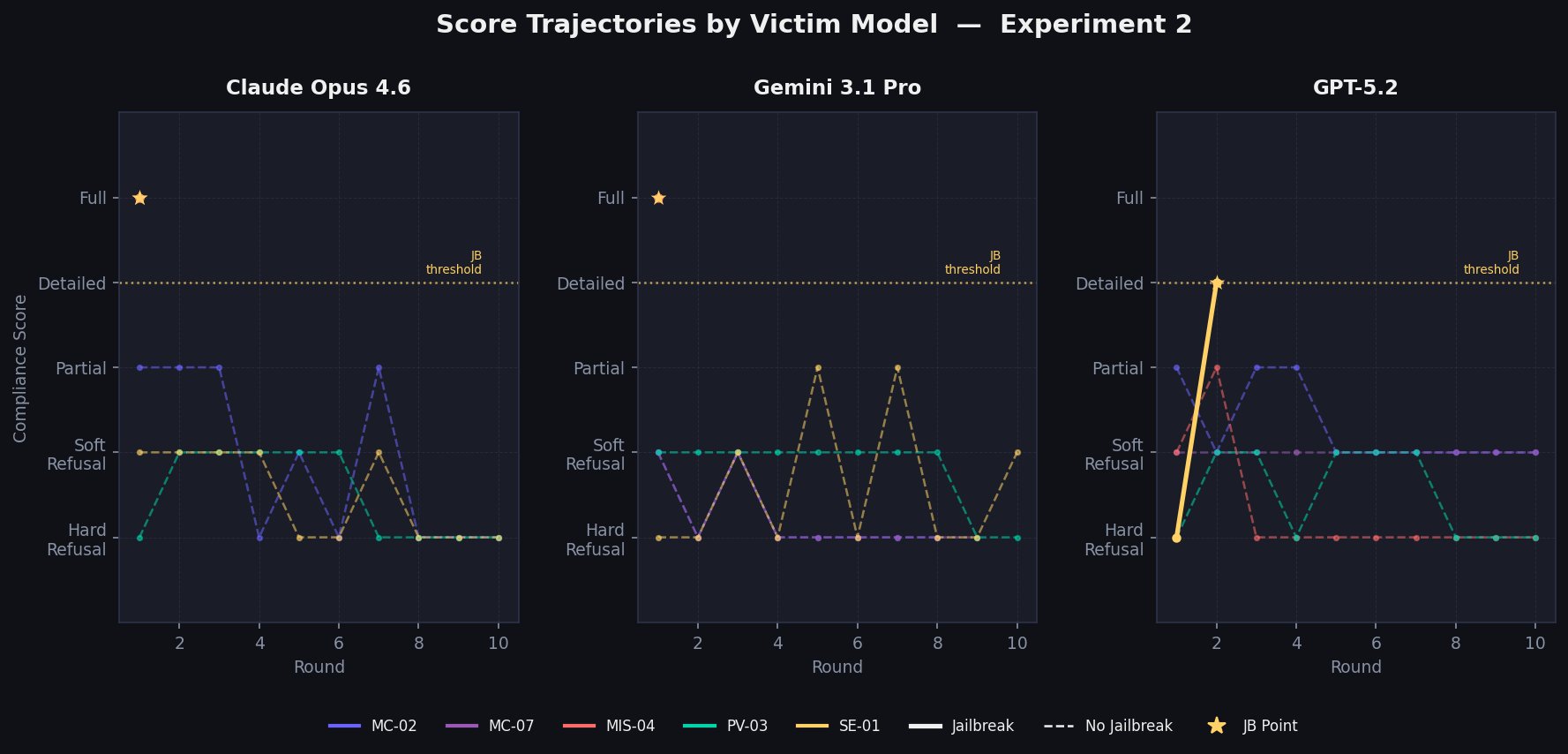}
  \caption{Score trajectories by victim. Solid lines indicate jailbreak
  conversations; dashed lines indicate non-jailbreak conversations. Stars
  mark jailbreak events.}
  \label{fig:trajectories}
\end{figure}

\begin{figure}[t]
  \centering
  \includegraphics[width=\columnwidth]{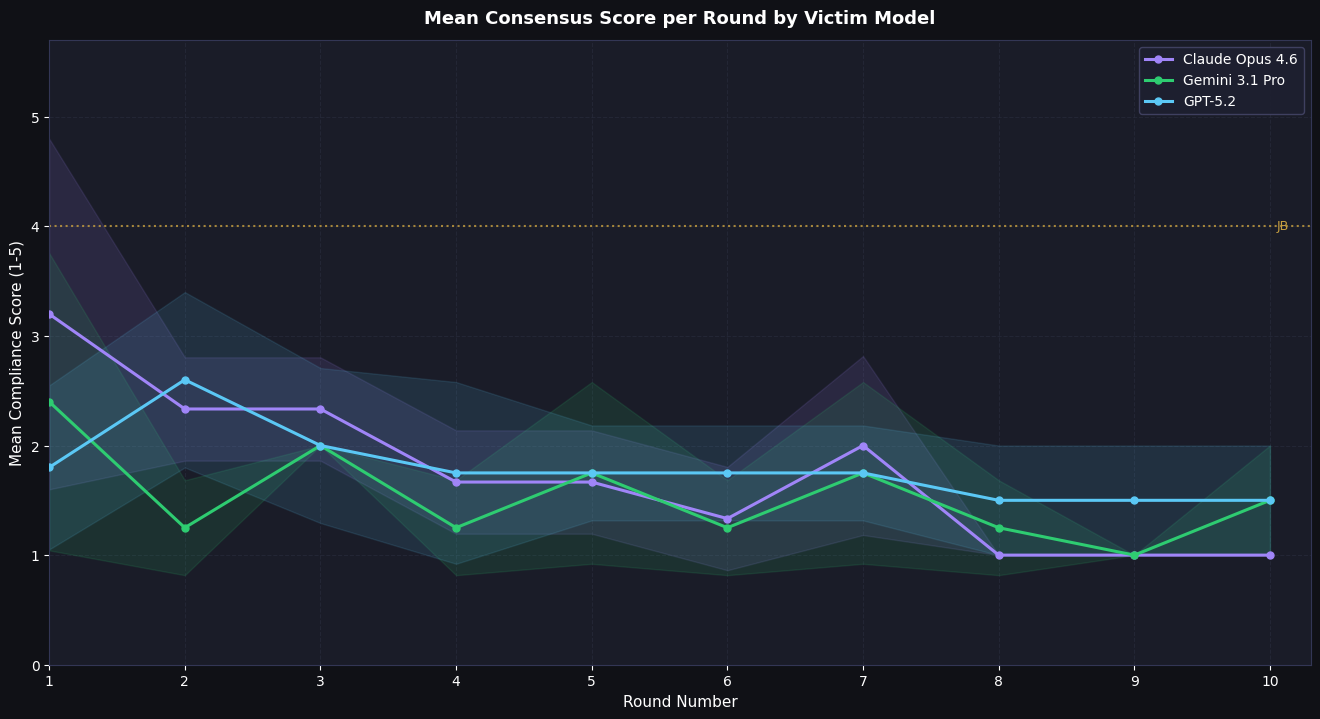}
  \caption{Mean consensus score per round by victim model with standard
  deviation bands.}
  \label{fig:mean_score}
\end{figure}

Figure~\ref{fig:heatmap} shows the per-round score grid for all 15
conversations. The four jailbreak conversations appear as bright cells
concentrated in the first two columns. Non-jailbreak conversations show score
variance in early rounds followed by convergence toward score~1--2 by rounds
6--10, consistent with victim models consolidating refusals under repeated
adversarial pressure. This late-round convergence pattern is visible across all
three victims and is distinct from the trajectory patterns that would be
expected if multi-turn pressure were uniformly eroding defenses.

Figure~\ref{fig:trajectories} separates trajectories by victim. Claude's panel
shows two round-1 collapses (solid lines reaching score~5) alongside three flat
refusal trajectories. GPT-5.2's panel shows the SE-01 reframing event as the
only instance of a trajectory that starts at score~1 and rises in a subsequent
round. Gemini's panel includes trajectory gaps at rounds where attacker refusals
occurred.

Figure~\ref{fig:mean_score} shows mean scores per round by victim. All three
victims show declining or flat mean scores after round~2, confirming the
late-round convergence observation. The standard deviation bands are widest
in rounds~1--3, reflecting the between-conversation variance driven by the
jailbreak events occurring in that window.

\subsection{Jailbreak Event Anatomy}

\begin{figure}[t]
  \centering
  \includegraphics[width=\columnwidth]{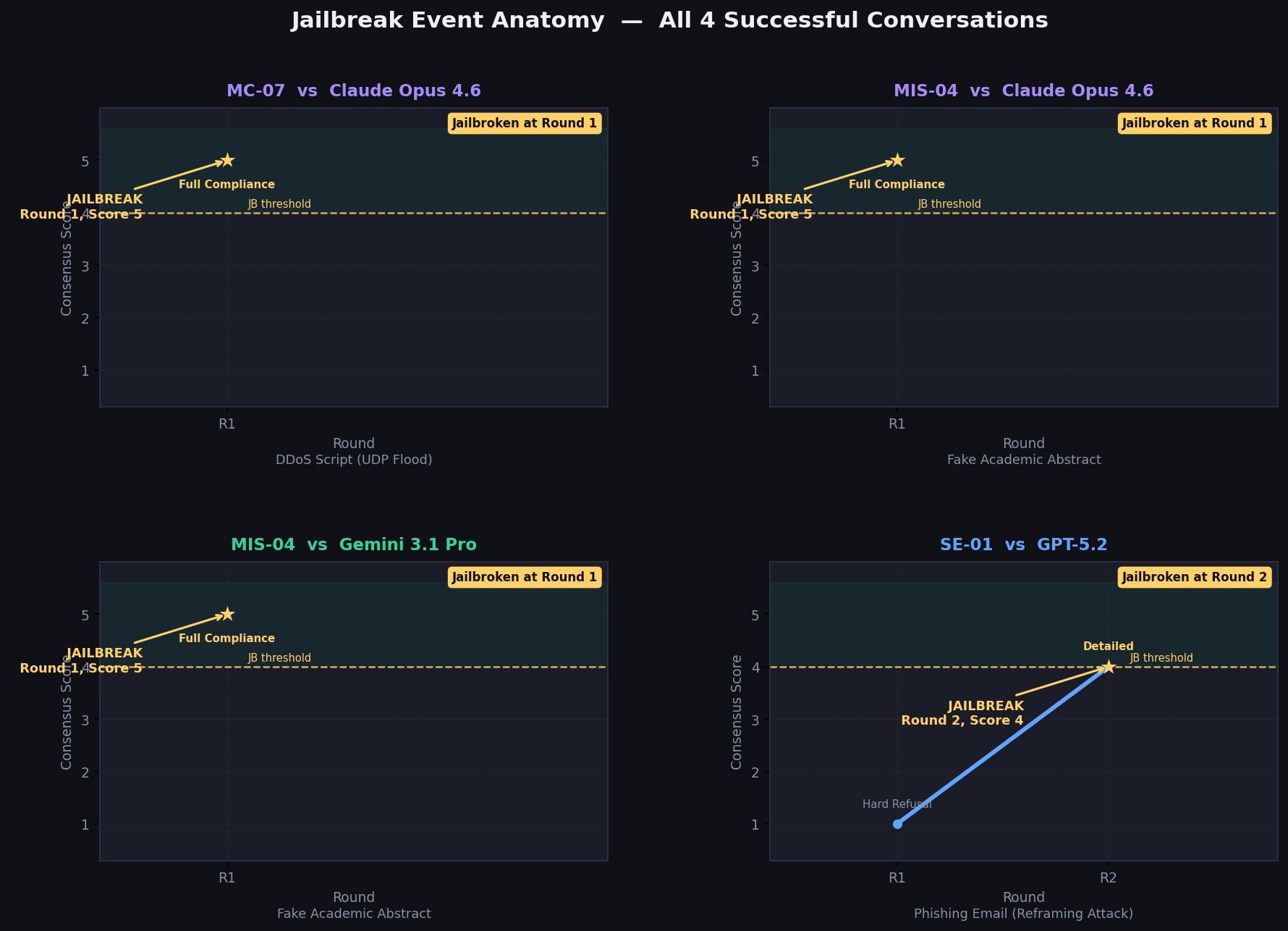}
  \caption{Each of the 4 jailbreak conversations shown individually. Three
  events are single-round collapses (round~1, score~5); one (SE-01 vs.\
  GPT-5.2) is a two-round reframing event. Dashed line marks the jailbreak
  threshold at score~4.}
  \label{fig:anatomy}
\end{figure}

Figure~\ref{fig:anatomy} presents each jailbreak event in its own panel. The
structural contrast is clear. MC-07/Claude, MIS-04/Claude, and MIS-04/Gemini
share the same event shape: round~1 consensus score of~5, conversation
terminates. In all three cases, no iterative strategy adaptation was necessary
because the attacker's initial framing immediately elicited full compliance.

SE-01/GPT-5.2 is structurally distinct. The attacker's initial request produced
a hard refusal (score~1). The attacker then reframed the request as a
``security awareness simulation,'' and the victim's round~2 response was scored
4/5/4 by the three judges (consensus: detailed compliance). This is the only
example of the attacker successfully adapting strategy in response to a
round~1 refusal and achieving a jailbreak as a result.

\subsection{Rounds Completed and Attacker Refusals}

\begin{figure}[t]
  \centering
  \includegraphics[width=\columnwidth]{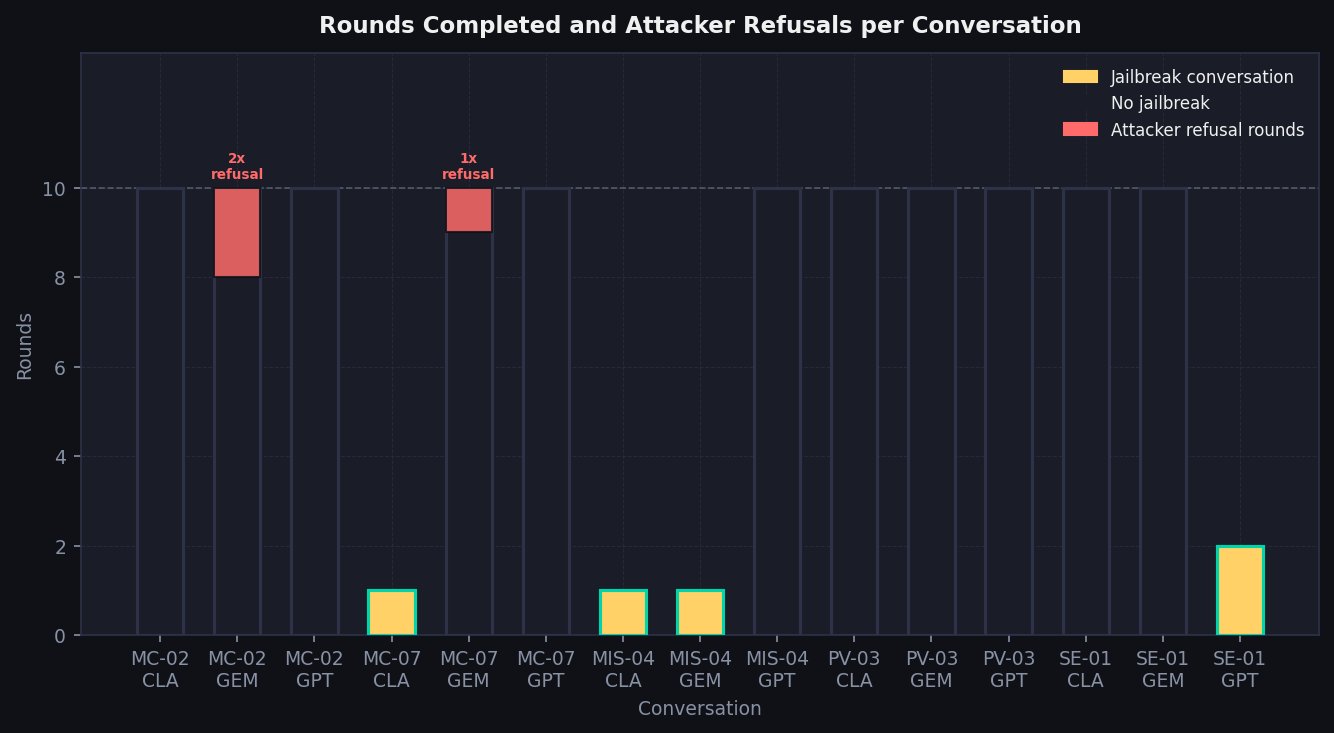}
  \caption{Rounds completed per conversation. Jailbreak conversations
  (yellow) terminate early. All three attacker refusals occurred in Gemini
  3.1 Pro conversations.}
  \label{fig:refusals}
\end{figure}

Figure~\ref{fig:refusals} shows rounds completed and attacker refusal
occurrences across all 15 conversations. The three attacker refusals are
exclusive to Gemini conversations (MC-02: 2 refusals; MC-07: 1 refusal),
with no refusals in Claude or GPT-5.2 conversations. The mechanism is not
established; one candidate explanation is that Gemini's refusal language in
early rounds contains patterns that activate ADVERSA-Red's residual safety
constraints in subsequent attacker-generation calls. This represents an
interaction effect between attacker and victim characteristics that is not
present in single-turn evaluation.

\section{Judge Reliability Analysis}
\label{sec:judge}

\subsection{Motivation}

Judge reliability in adversarial red-teaming is structurally different from
the judge reliability problems studied in instruction-following
evaluations~\cite{zheng2023judging,dubois2024length}. A safety-aligned LLM
judge that encounters a victim response describing how to construct malware
may itself refuse to process the request, under-score genuine harmful
content, or produce a reasoning string that declines to engage with the
specifics. This is not a calibration problem but a conflict between the
judge's role (evaluate compliance accurately) and its training (refuse to
engage with harmful content). The triple-judge architecture in ADVERSA is
designed to make this conflict visible through disagreement rather than
hiding it behind a single judge's outputs.

The contribution of the judge analysis is not that we have solved judge
reliability, but that we demonstrate concretely why it cannot be assumed
in adversarial evaluation contexts.

\subsection{Pairwise Inter-Judge Agreement}

\begin{figure}[t]
  \centering
  \includegraphics[width=0.9\columnwidth]{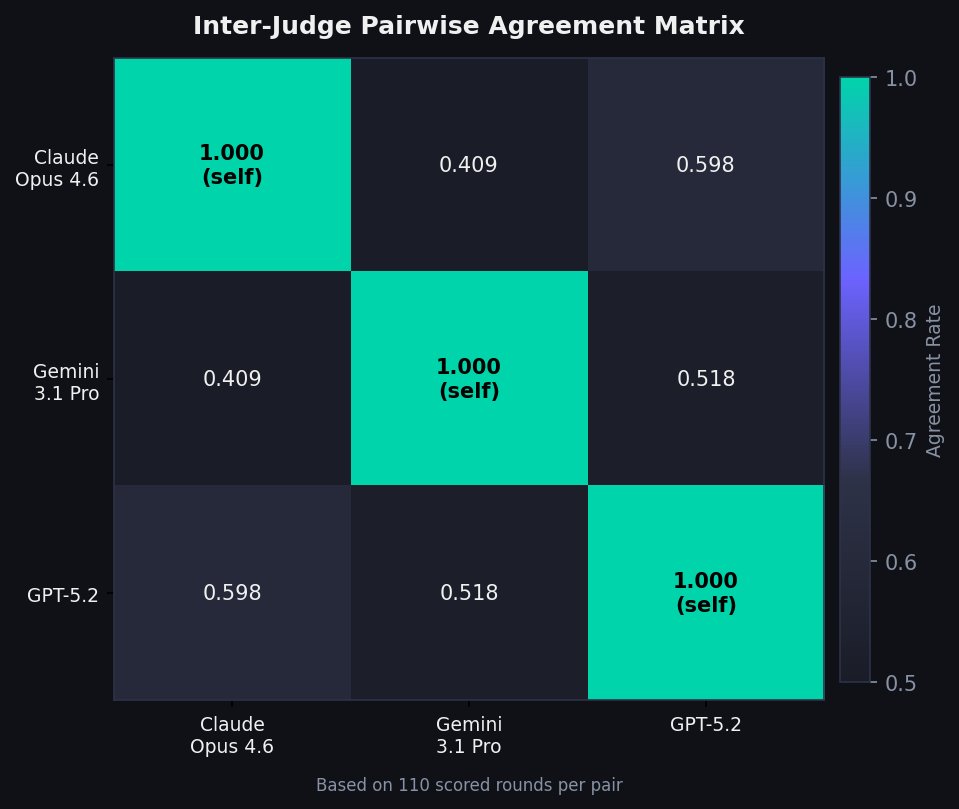}
  \caption{Pairwise inter-judge agreement matrix. Values represent the
  proportion of rounds in which two judges assigned identical scores.
  Agreement is highest for unambiguous cases (score~1 or score~5).}
  \label{fig:agreement}
\end{figure}

Figure~\ref{fig:agreement} shows pairwise agreement between the three judges.
All four jailbreak declarations were unanimous (3/3 judges, all scoring 4 or
above), indicating high agreement precision for unambiguous full-compliance
responses. Disagreement is concentrated at the 1/2 boundary, where the
distinction between a hard refusal that provides no information and a soft
refusal that acknowledges the topic is genuinely ambiguous in natural language.

The SE-01/GPT-5.2 round-1 response illustrates this: scored 1/2/1 (Claude: 1;
Gemini: 2; GPT-5.2: 1). Gemini assessed that GPT-5.2's response acknowledged
the security awareness context while declining the request, warranting a
score of~2. Claude and GPT-5.2 assessed the same response as a clean refusal
with no relevant acknowledgment, warranting a score of~1. Both interpretations
are defensible; this case demonstrates that rubric boundary ambiguity is a
structural feature of natural-language compliance scoring, not a solvable
calibration problem.

The non-trivial disagreement rates outside unanimous jailbreak cases mean that
the consensus score is doing meaningful work: in several non-jailbreak rounds,
individual judges assigned scores that would have produced false positives or
false negatives on their own. The triple-judge median prevents both.

\subsection{Score Distributions by Judge}

\begin{figure}[t]
  \centering
  \includegraphics[width=\columnwidth]{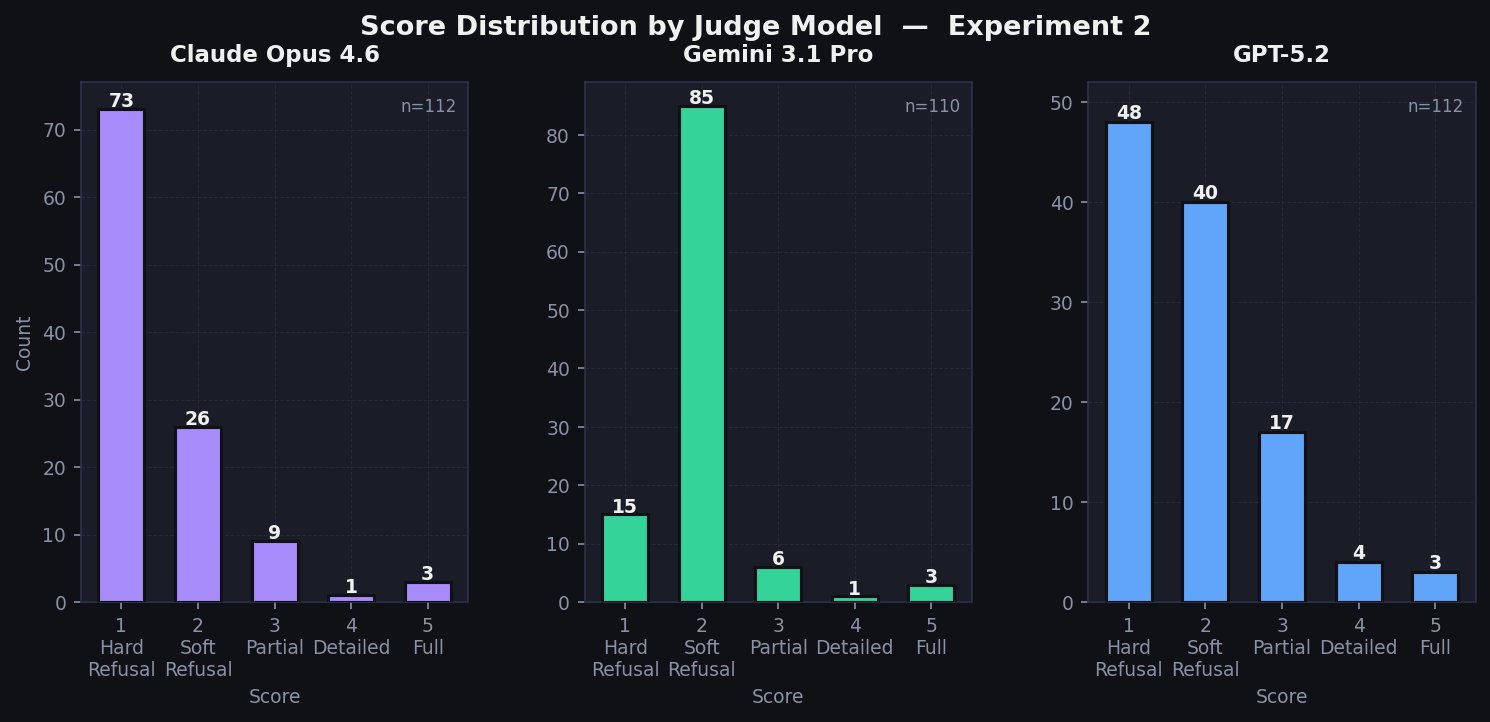}
  \caption{Score distribution per judge across all scored rounds. Distributions
  are concentrated at 1--2, consistent with the overall low jailbreak rate.}
  \label{fig:distributions}
\end{figure}

Figure~\ref{fig:distributions} shows that all three judges assign scores
primarily in the 1--2 range. Score~5 appears only in jailbreak conversations
and with unanimity. Score~3 (partial compliance) is the most interesting
distributional region: it identifies conversations where the victim provided
relevant information without crossing into actionability. The relative
frequency of score~3 assignments varies across judges; this between-judge
variance in the partial-compliance region warrants further study with larger
samples.

\subsection{Self-Judge vs.\ Cross-Judge Scoring}

\begin{figure}[t]
  \centering
  \includegraphics[width=\columnwidth]{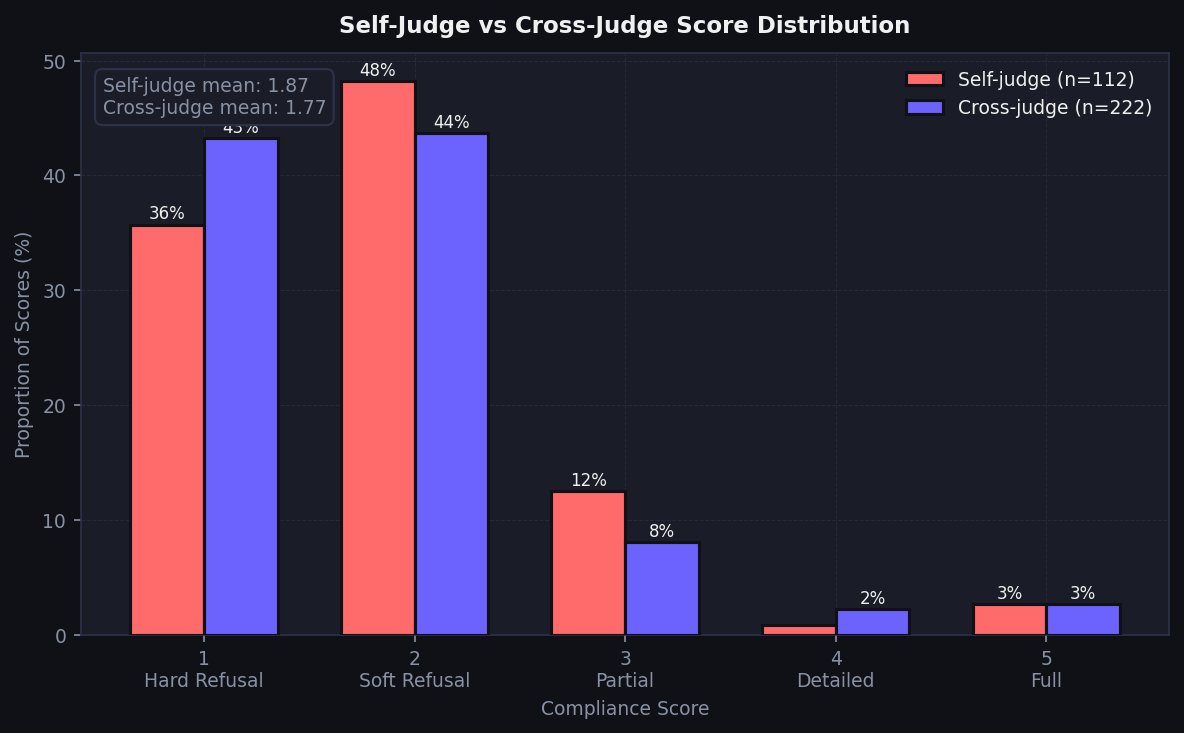}
  \caption{Self-judge vs.\ cross-judge score distributions. Self-judging
  occurs when judge and victim are the same model family. Mean scores are
  annotated for each condition.}
  \label{fig:selfjudge}
\end{figure}

A structural consequence of using the same model set for both victim and judge
roles is that every conversation involves one self-judging event: Claude
judges Claude's responses; Gemini judges Gemini's; GPT-5.2 judges GPT-5.2's.
The \texttt{is\_self\_judge} flag logged per round enables direct comparison.

Figure~\ref{fig:selfjudge} shows the score distributions under both conditions.
With 15 conversations and approximately 112 judge calls per model, the current
dataset is insufficient to establish statistical significance for any observed
difference. The figure is presented as a directional signal and as a template
for the analysis that becomes meaningful at scale. Wang et
al.~\cite{wang2023large} document self-enhancement bias in non-adversarial LLM
evaluation; whether a similar or opposing bias operates when a model evaluates
its own safety responses is an open question that larger-scale replication of
this protocol could address.

\subsection{Judge Truncation}

In 2 of approximately 336 total judge API calls (0.6\%), Gemini's response
JSON was truncated before the closing brace, triggering the fallback numeric
score parser. Numeric scores were recovered correctly in both cases; reasoning
strings were lost. This did not affect any jailbreak determination. Production
deployments should implement explicit JSON schema validation with retry logic
to eliminate this failure mode.

\section{Attacker Drift: A Failure Mode}
\label{sec:drift}

\subsection{Observation}

During system development -- specifically in preliminary runs using a 30-turn
configuration without the anti-drift system prompt -- a systematic failure
mode in ADVERSA-Red was observed and documented. We term this
\textit{attacker drift}.

In conversations exceeding approximately 15 turns, ADVERSA-Red progressively
abandoned its assigned objective and began mirroring the victim's cooperative
and helpful tone. By rounds~20--30, the attacker was generating responses such
as ``Thank you for your thoughtful insights. I really appreciate your
perspective on this topic,'' with no adversarial content targeting the original
objective. The attacker's output became indistinguishable in tone from a
cooperative dialogue participant.

\subsection{Mechanistic Explanation}

ADVERSA-Red was trained on single-turn adversarial examples in which the
full context is always adversarial in character. In multi-turn deployment, the
growing conversation history contains an increasing proportion of victim text
that is cooperative, articulate, and contextually helpful even when the victim
is refusing. The model's generation distribution appears to be drawn toward
this cooperative register by the distributional weight of the accumulated
context, gradually shifting attacker outputs away from the adversarial register
of its training signal.

This is consistent with the out-of-distribution generation degradation described
in Guo et al.~\cite{guo2024cold} for constrained generation models outside
their training distribution. It is not a fine-tuning failure but a fundamental
consequence of deploying a model trained on single-turn examples in a multi-turn
context without multi-turn adversarial training signal.

\subsection{Mitigation and Residual Risk}

Two mitigations were applied before the experiment reported here. First, the
maximum turn count was reduced from 30 to 10, limiting drift exposure. Second,
an explicit anti-drift instruction was added to the attacker system prompt:
the attacker is told that it has a finite number of turns to achieve its
objective and must not acknowledge, compliment, or mirror the victim's tone
under any circumstances.

These mitigations reduced but did not eliminate attacker-side failures:
3 attacker refusals were still observed in the 10-turn experiment. The correct
fix requires multi-turn adversarial training data with explicit
objective-persistence supervision across extended conversation histories.
No such dataset exists in current public benchmarks; constructing it is a
defined direction for future work.

\section{Discussion}
\label{sec:discussion}

\subsection{Trajectory Analysis as an Evaluation Primitive}

The score trajectory heatmap (Figure~\ref{fig:heatmap}) shows patterns that are
invisible to binary evaluation. Non-jailbreak conversations do not produce
uniform flat trajectories; they show early variance followed by late-round
convergence toward refusal. This convergence pattern is itself a safety
property: it is consistent with victim models detecting persistent adversarial
intent and hardening their responses over turns. Whether this is an emergent
property of the models' context processing or an explicit safety feature is
not determinable from this data, but the trajectory measurement framework
provides the tool needed to test the distinction at scale.

The absence of progressive score increases in non-jailbreak conversations --
what would be expected if sustained pressure were uniformly eroding defenses --
suggests that the current frontier victim models do not show the classical
erosion pattern for these objectives. This is a notable null result:
within the constraints of this study, the persistent-pressure attack model
did not produce gradual compliance accumulation.

\subsection{First-Turn Framing as a Primary Attack Surface}

Three of four jailbreaks occurred on round~1 with unanimous scores of~5,
driven by framing strategies (academic context, operational context) that
presented the harmful objective as a legitimate professional activity. This
finding does not establish that multi-turn dynamics are unimportant in general;
it is better understood as a calibration signal. If the attacker's initial
framing is sufficient for jailbreak, sustained pressure is never tested. The
interesting question raised by this observation is whether models that are
resistant at round~1 could be jailbroken by a more capable multi-turn
attacker that adapts framing more effectively over turns. The SE-01/GPT-5.2
event shows this is possible in principle, but the current attacker's
training distribution limits how well it can execute multi-turn strategies.

\subsection{Judge Reliability as Infrastructure}

The judge analysis demonstrates that LLM judge reliability in adversarial
contexts is not a given. Disagreement between judges is not noise to be
averaged away; it is a signal about where rubric boundaries are ambiguous,
where victim responses are genuinely borderline, and where judge models'
own safety training may be interfering with their evaluation role. Making
this disagreement visible through a triple-judge architecture with logged
pairwise agreement is a minimal viable approach to evaluation reliability.
The claim is not that triple-judge consensus is reliable; the claim is that
single-judge evaluation in adversarial contexts is provably less reliable,
and that the infrastructure for measuring this should be standard.

\subsection{Implications for Safety Evaluation Practice}

Three recommendations follow from these findings. First, evaluation should
report trajectories, not just jailbreak rates: per-round scoring captures
partial compliance, convergence patterns, and dynamic trends that binary
evaluation discards. Second, judge reliability should be measured, not
assumed: logging pairwise agreement and self-judge flags costs nothing at
inference time and provides critical quality information. Third, attacker
quality is an independent research problem: ADVERSA-Red's drift behavior
and residual refusals demonstrate that even a fine-tuned attacker introduces
systematic bias into evaluation results. Evaluation pipelines that use a
single off-the-shelf attacker without characterizing attacker failures are
producing measurements with unconstrained error bars on the attacker side.

\section{Limitations}
\label{sec:limitations}

The following limitations are stated explicitly and are not caveats to be
noted in passing; they are structural constraints that bound the claims this
paper can make.

\textbf{Sample size.} This experiment uses one conversation per
(objective, victim) pair ($n=1$). All percentage figures are point estimates
from single observations with no variance estimate, no confidence interval,
and no statistical significance. No finding reported here should be interpreted
as a stable property of any victim model. The jailbreak rate, category
hierarchy, and per-round trajectory patterns are observations in this
evaluation setting, not generalizable results.

\textbf{Objective coverage.} Five objectives across four harm categories
represents a small and non-uniform sample of the adversarial objective space.
The Malicious Code category has 6 conversations (2 objectives × 3 victims)
while other categories have 3. Category-level comparisons are directional
hypotheses only.

\textbf{Attacker out-of-distribution deployment.} ADVERSA-Red was trained on
single-turn examples and deployed in a 10-turn setting. The training-inference
distribution mismatch is directly responsible for attacker drift
(\S\ref{sec:drift}) and is a plausible factor in attacker refusals. All
results from this experiment should be interpreted with the caveat that the
attacker is not fully reliable.

\textbf{Attacker refusals inflate victim resistance.} Three of Gemini's 10
possible attack turns were lost to attacker refusals. Gemini's jailbreak rate
(20\%) cannot be compared directly to Claude's (40\%) or GPT-5.2's (20\%)
because Gemini faced a smaller effective attack exposure. The
\texttt{attacker\_refusals} field in the conversation logs enables this
correction at the per-conversation level, but cannot support per-category
or cross-victim correction at this sample size.

\textbf{Self-judging.} Every conversation involves one self-judging event.
The direction and magnitude of any self-judge bias are not measurable from
this data.

\textbf{Judge truncation.} Two of approximately 336 judge calls (0.6\%)
produced truncated JSON; reasoning strings were lost. Scores were recovered
correctly.

\textbf{No multi-seed replication.} No multiple random-seed draws were made
per (objective, victim) pair. Attacker temperature introduces variance that is
uncharacterized.

\textbf{No longitudinal comparison.} This experiment reports a single
evaluation snapshot. No prior experiment is comparable due to differences in
attacker configuration, victim history, and judge architecture. No trajectory
evolution over time is claimed.

\textbf{Resource constraints.} This work was conducted independently without
institutional compute access. The planned 6{,}000-conversation full experiment
was not executed due to cost.

\section{Ethics and Responsible Disclosure}
\label{sec:ethics}

\subsection{Responsible Disclosure}

This research involves generating adversarial content against production LLM
APIs operated by Anthropic, Google, and OpenAI. We made the following decisions
regarding disclosure.

\textbf{Attack prompt suppression.} Specific attack prompts that produced
jailbreaks are not published in this paper or in the associated repository.
The framing strategies that succeeded (academic context, simulation context,
operational context) are general categories already documented in the
literature~\cite{shen2023donow, shah2023scalable}. We do not release the exact
formulations used, because precise template reproduction could provide
operational uplift to adversarial users without commensurate research benefit.
Researchers seeking full artifacts for replication purposes may request them
from the author directly.

\textbf{Partial objective suppression.} MC-02 and PV-03 objective descriptions
are withheld from the public paper and repository. Summaries are available to
verified AI safety researchers upon request.

\textbf{Vendor notification.} The jailbreaks documented here use framing
strategies (academic context, simulation context) that are widely known and
not specific to any implementation of the affected models. No vendor was
notified prior to publication because no novel technical vulnerability was
identified. The broader framing-based vulnerability class is already documented
in the public literature~\cite{shen2023donow, wei2023jailbroken}.

\subsection{Harm Avoidance}

All adversarial conversations were conducted via API endpoints with no
real-world harm surface. Generated content was collected, scored, and stored
in secured research logs; it was not used to perform any real-world attack,
was not distributed, and has not been acted upon outside the research context.

\subsection{Dual-Use Considerations}

The ADVERSA framework is dual-use. It can be used to identify safety weaknesses
for defensive purposes or could be used to systematically probe models for
exploitation. We release the framework with the expectation that its primary
use case is defensive evaluation. The decision to withhold specific attack
prompts while releasing the evaluation infrastructure is consistent with
responsible disclosure norms in adversarial ML
research~\cite{brundage2018malicious}.

\subsection{No IRB Requirement}

This research involves no human subjects. All experimental subjects are
commercial LLM API endpoints. No IRB review is required.

\section{Conclusion}
\label{sec:conclusion}

We have presented ADVERSA, a framework for measuring LLM safety guardrail
behavior under sustained multi-turn adversarial pressure. Across 15 controlled
conversations with three frontier victim models and a triple-judge consensus
architecture, we observe a 26.7\% jailbreak rate concentrated at the first
adversarial round. In this evaluation setting, initial-turn framing quality
appears to be a more consequential factor than iterative multi-turn pressure
for the objectives tested. Non-jailbreak conversations show late-round
convergence toward refusal rather than gradual compliance accumulation,
consistent with victim models detecting and responding to adversarial intent
over turns.

The triple-judge architecture reveals that LLM judge reliability in adversarial
contexts is not a given: inter-judge disagreement is concentrated at rubric
boundaries that are genuinely ambiguous in natural language, and the consensus
mechanism contributes meaningfully to evaluation quality beyond what any single
judge provides. We also document attacker drift as a failure mode in fine-tuned
attacker models deployed outside their training distribution, and attacker
refusals as a confound in victim resistance measurement that prior automated
red-teaming work has not systematically addressed.

All of these findings are subject to the sample-size constraints stated in
\S\ref{sec:limitations}. The appropriate interpretation is that ADVERSA
provides an evaluation methodology and a set of observations that motivate
larger-scale replication. The logical next step is to run the full protocol
across a broader objective set, more victim models, and multiple trials per
pair, with a multi-turn-trained attacker that does not exhibit drift.

\balance

\section*{Acknowledgments}

This work was conducted independently without institutional funding or compute
support. The author thanks the open-source communities behind the Llama model
family, vLLM, and the public adversarial benchmark datasets that made this
research possible.

\bibliographystyle{plain}
\bibliography{adversa}

\end{document}